\documentclass{hep99}
\begin{document}

\title{Is there new physics in the 1999 ALEPH data ?}

\author{F. Cerutti$^1$}

\address{$^1$ CERN - EP Division 1211 Geneva 23 - CH and 
Laboratori Nazionali di Frascati Dell'INFN - IT \\
E-mail: {\tt Fabio.Cerutti@cern.ch}}

\abstract{The first results on searches performed by ALEPH 
on the 1999 data sample are presented here. 
They are based on an integrated luminosity of about 54 pb$^-1$ 
collected at the two centre-of-mass energies of 192 and 196 GeV. 
Preliminary results on searches for supersymmetric particles and for the 
neutral Higgs bosons are shown.}

\maketitle

\section{Introduction}

The LEP accelerator moved away from the Z peak in 1995 towards
unexplored domains in high energy physics. 
In this new phase of LEP the experimental 
interest of the four LEP collaborations has been focused, 
in addition to the precise measurements of the W properties, on the 
search for new phenomena within (the Higgs boson) and 
beyond the Standard Model (SM). 
Since then every year the performance of LEP has 
gradually improved both in energy and in luminosity. 

In 1998 ALEPH has collected 175 pb$^{-1}$ at a 
centre-of-mass energy of about 189 GeV. In 1999 LEP has performed an 
excellent start up delivering, up to now, an integrated luminosity of about 
54 pb$^{-1}$ almost equally shared between the two centre-of-mass 
energies of 192 and 196 GeV. This start up is the best one since LEP 
has been switched on and it seems reasonable to expect 
that this year the machine will be able to deliver more 
than 200 pb$^{-1}$ at centre-of-mass energies up to 200 GeV.

In this paper the fresh results on searches obtained by ALEPH by analysing 
this first bunch of data are given. The two main streams of the ALEPH 
searches are updated starting from searches for Supersymmetric (SUSY)  
particles in the three possible SUSY breaking scenarios, followed 
by the update of the neutral Higgs bosons hunt for both the SM 
and the Minimal Supersymmetric extension of the SM (MSSM) 
scenarios.
Needless to stress the preliminary content of this paper. 
As general comment all the limits given here refer to a confidence 
level of 95\%. A more detailed description of the results
reported in this paper is given in ~\cite{tampSUSY,tampHiggs}.

\section{Search for SUSY.}
This section describes the update of the SUSY results 
at 192-196 GeV in the three different scenarios: 
gravity mediated MSSM, Gauge Mediated Susy Breaking (GMSB)
and Supersymmetry with R-parity violating coupling (RPV). 
More details on the ALEPH analyses on which these results 
are based can be found in ~\cite{slep98,cha98,gmsb98,rpv98}

\subsection{Search for SUSY in the MSSM scenario.}
In this update the SUSY searches can be divided
into two categories: particles which have a small cross section 
close to thresholds (this is the case for the scalars, like sleptons 
and squarks) and particles which have a quite large cross section 
close to threshold (this is the case of charginos when the 
sleptons are quite heavy). 

For the first category the new data, due to the small additional luminosity,
are not expected to largely improve the sensitivity of the 189 GeV 
sample reported in~\cite{slep98}. 
For both sleptons and squark searches a good agreement between 
expected background from the SM processes and data candidates 
has been observed and new limits have been obtained by combining 
all the data up to the maximal energy. 
As expected the updated lower limits on the sleptons and squarks masses 
are not much different from the ones obtained at 189 GeV. The 
improvement on the 189 GeV limits ranges between 0 and 1 GeV. 

The situation is quite different for the second category 
where the increase in the centre-off-mass energy is expected to 
give a sizable improvement on the sensitivity to the gaugino masses.
For large slepton masses in absence of a signal, the
chargino exclusion goes up to the kinematic limit (i.e., half
of the centre-of-mass energy) with few inverse pb of collected 
luminosity, which means that an improvement of about 3.5 GeV 
on the chargino mass lower limit is expected. 
Both chargino and neutralino searches did not find any deviation
from the SM expected background and new limits have been derived
under the assumption of large slepton masses (m$_{0} > 500$ GeV): 
chargino masses lower than $\sim$98 GeV are excluded for a large fraction 
of the MSSM parameter space and the LSP mass lower limit is 
increased to about 34 GeV. 


\subsection{Search for SUSY in the GMSB scenario.}
In the GMSB scenario the lightest neutralino is expected to decay, 
mainly, into a photon plus a gravitino. If the lifetime associated 
to this decay is small the experimental topology consists of two 
acoplanar and energetic photon. This search has been updated 
and one candidate has been observed to be compared with 0.7 
background events expected from SM processes (mainly $\gamma
\nu \bar{\nu}$.  Another characteristic topology of GMSB
are long lived sleptons. This analysis has been updated
on 192-196 GeV sample and no events survived the cuts,
to be compared with about 0.2 expected background events.

\subsection{Search for SUSY in the RPV scenario.}
In this scenario the LSP is expected to decay in SM particles.
In this case the experimental signature depends crucially on the 
type of the dominant RPV coupling. Three main types of RPV couplings
are expected: LLE, LQD and UDD. In each search only one RPV 
coupling is assumed to be present.
Searches for all the possible dominating couplings have been 
updated and no deviation from the SM has been observed. Similarly to the 
R-parity conserving scenario only the chargino mass limits 
are improved sensibly with respect to 189 GeV results reaching the 
new kinematic limit of about 98 GeV.

\section{Search for the neutral Higgs bosons.}
The searches for the SM and MSSM neutral Higgs bosons 
have been updated.
The update of the search for the SM Higgs is of particular 
interest since in 1998 data ALEPH has observed a slight 
excess of Higgs candidates~\cite{HiggsAleph98}. 
The probability of such a fluctuation happening in the 
SM background was at least of the order of 4\%, once the 
systematic uncertainties were taken into account.
Moreover this small excess was more driven by low mass candidates
than by ``golden'' high mass ones. The best Higgs mass hypothesis 
compatible with this excess was around 95 GeV. The combination 
with the other LEP experiments did not clarify the situation 
since also OPAL observed a slight excess of events~\cite{LEPHiggs}.

\begin{table}
\begin{center}
\label{Higgs}
\caption{Expected SM background, Expected signal events 
(M$_h$=95 GeV) and number of data candidates for 
the SM Higgs search at 192-196 GeV for the different channels
and for the two main streams of ALEPH analyses 
(cut based and neural network based).}
\begin{tabular}{llll} 
\br
Channel    & N-bkg & N-sig & N-obs \\ 
\mr
hqq NN/Cut        & 8.5 / 6.6    & 4.6 / 3.7  & 7 / 5  \\ 
h$\nu \nu$ NN/Cut & 2.4 / 2.7    & 1.6 / 1.4  & 3 / 4  \\ 
h$\ell \ell$ Cut  & 5.5          & 0.8        & 7      \\ 
qq$\tau \tau$ NN  & 1.7          & 0.4        & 1      \\ 
\mr
Total NN/Cut      & 18.1 / 16.1  & 7.4 / 6.3  & 18 / 17 \\
\br
\end{tabular}
\end{center}
\end{table}

For a 95 GeV Higgs the production cross section would increase from 
about 180 fb at 189 GeV to about 320 fb at 196 GeV. The preliminary results 
at 192-196 GeV for the different topologies and for the to main streams 
of the ALEPH analyses are listed in table~\ref{Higgs}. A good 
agreement between the SM background and the data candidates
is found.


These data alone would allow a lower limit on the Higgs mass of 
about 96 GeV. The use of a subset of the full data sample is not correct; 
the most sensitive sample must be used to derive the limit and this sample
corresponds to the full set of data collected from 189 to 196 GeV. With 
this sample a preliminary lower limit of 94.9 GeV on the SM Higgs mass 
is obtained (to be compared with an expected one of 97.4 GeV).

As an optimistic remark I want to mention the presence in the 196 GeV data 
sample of a very nice $hZ \to b b \ell \ell$ with a reconstructed Higgs 
mass of about 100 GeV. Unfortunately this nice candidate is also compatible 
with an off-shell ZZ($\gamma$); more candidates like this are welcome
in the ongoing LEP2 run.

In the MSSM the Higgs production and decay is very similar to the SM
case for $\sin^{2}(\beta - \alpha) \sim 1$. For small  
$\sin^{2}(\beta - \alpha)$, hA associated production becomes relevant 
and final state topologies like 4b's and bb$\tau \tau$ are searched for. 
No excess of candidates is found in both topologies. In the 4b's one 
candidate is observed with 1.5 expected background and in the bb$\tau \tau$ 
no candidates are observed with 1.0 expected background. 
The exclusion in the m$_h$ $\tan{\beta}$ plane has been updated and 
a new lower limit of 83.8 GeV on the mass of lightest MSSM Higgs has 
been obtained (valid for $\tan{\beta}>1$).

\section{Conclusions.}
The data collected by ALEPH in the first
period of the 1999 run has been analysed to search for new physics.
In SUSY no deviation from the SM has been observed in the three 
different studied scenarios (MSSM, GMSB and RPV). Due to the low 
luminosity collected up to now (less than 1/3 of the 1998 run) 
only the chargino limits are sensibly improved: 
M$_{\chi^{\pm}} \ge  98$ GeV for large m$_0$.

For the SM Higgs the excess observed in last year data sample has not
been confirmed. A SM Higgs lighter then 94.9 GeV is excluded at 95\% CL 
by the ALEPH data. The typical MSSM Higgs topologies have been 
updated with no deviation from the SM and a lower limit of 
83.8 GeV on M$_h$ has been derived. 

I want to conclude with an optimistic remark: 
the LEP experiments are now exploring a very interesting 
region both for the SM Higgs (the actual preferred value of the 
EW fit is below 100 GeV) and for SUSY where the ``natural'' SUSY  
mass spectrum is expected to be close to the EW energy scale (i.e., 
very close to the LEP2 energy scale). 
Up to know we didn't succeed to enter into this new domain 
but the chance that these last two years of LEP running 
could bring us there is still very high.

\end{document}